\documentclass[aps,prd,twocolumn,superscriptaddress,nofootinbib]{revtex4}

\usepackage{graphicx}
\usepackage{amsmath,amssymb,latexsym}
\usepackage{bm}
\usepackage{epsfig}

\newcommand{\met}{\mbox{$\not\!\!E_{T}$}}

\newcommand{\postscript}[2]{\setlength{\epsfxsize}{#2\hsize}
   \centerline{\epsfbox{#1}}}

\begin{document}

\def\rhadron.{$R$-hadron}
\def\rhadrons.{$R$-hadrons}

\hspace*{135mm} {\large \tt CERN-PH-TH/2007-120}

\title{Hunting long-lived gluinos at the Pierre Auger Observatory}

\author{Luis A.~Anchordoqui}
\affiliation{Department of Physics, University of Wisconsin-Milwaukee,
P.O. Box 413, Milwaukee, WI 53201, USA
}

\author{Antonio Delgado}
\affiliation{CERN, Theory Division, CH-1211 Geneva 23, Switzerland
}
\affiliation{Department of Physics, University of Notre Dame, 
Notre Dame, IN 46556, USA}

\author{Carlos A. Garc\'{\i}a Canal}
\affiliation{Departamento de F\'{\i}sica Universidad Nacional de 
La Plata, C.C.67, La Plata (1900), Argentina}
\affiliation{IFLP (CONICET), Universidad Nacional de La Plata, C.C.67, 
La Plata 1900, Argentina}

\author{Sergio J. Sciutto}
\affiliation{Departamento de F\'{\i}sica Universidad Nacional de 
La Plata, C.C.67, La Plata (1900), Argentina}
\affiliation{IFLP (CONICET), Universidad Nacional de La Plata, C.C.67, 
La Plata 1900, Argentina}

\date{October 2007}
\begin{abstract}
  \noindent
  Eventual signals of split sypersymmetry in cosmic ray physics are
  analyzed in detail. The study focusses particularly on quasi-stable
  colorless $R$-hadrons originating through confinement of long-lived
  gluinos (with quarks, anti-quarks, and gluons) produced in $pp$
  collisions at astrophysical sources. Because of parton density
  requirements, the gluino has a momentum which is considerable
  smaller than the energy of the primary proton, and so production of
  heavy (mass $\sim 500~{\rm GeV}$) $R$-hadrons requires powerful
  cosmic ray engines able to accelerate particles up to extreme
  energies, somewhat above $10^{13.6}~{\rm GeV}.$ Using a realistic
  Monte Carlo simulation with the AIRES engine, we study the main
  characteristics of the air showers triggered when one of these exotic
  hadrons impinges on a stationary nucleon of the Earth atmosphere. We
  show that $R$-hadron air showers present clear differences with
  respect to those initiated by standard particles.  We use this
  shower characteristics to construct observables which may be used to
  distinguish long-lived gluinos at the Pierre Auger Observatory.
\end{abstract}


\maketitle

\section{Introduction}

There exists ``lore'' that convinces us that physics beyond Standard
Model (SM) should be guided from the stabilization of mass hierarchy.
The most ubiquitous example is the minimal low energy effective
supersymmetric theory (MSSM)~\cite{Dimopoulos:1981zb}, which requires
a scale of supersymmetry (SUSY) breaking $\Lambda_{\rm SUSY} \sim
1$~TeV to avoid the fine tuning problem ($f_{\rm lore} \sim
M^2_H/\Lambda_{\rm SUSY}^2$) with the Higgs mass ($M_H \sim 100~{\rm
GeV}$). However, this ``naturalness'' is not favored by prescision
tests at colliders, which are consistent with  SM to a great
accuracy~\cite{Hagiwara:pw}. Consequently, any new physics which may
turn on beyond the electroweak scale needs to be
fine tuned at the percent level. Moreover, the presence~\cite{Bahcall:1999xn} of a tiny, but
non-vanishing, cosmological constant presents us with a fine tuning problem
much more severe than the gauge hierarchy problem.

The solution of last resort to address the cosmological constant
problem is Weinberg's anthropic approach~\cite{Weinberg:1987dv},
in which there exists an enormous ``landscape'' of vacua, only a
small fraction of which have a vacuum energy small enough to allow
for a natural habitat for observers such as ourselves. This
approach has been recently rekindled by investigations in String
Theory which have applied a statistical analysis to the large
number $N$ of vacua in the theory~\cite{Bousso:2000xa}. Among this
vast number of metastable vacua, there can be a small subset
${\cal O} (10^{40})$ exhibiting low scale SUSY breaking.  Of
course the fine tuning required to achieve a small cosmological
constant implies the need of a huge number of vacua, far more than
the ${\cal O} (10^{40})$ characterizing low-scale SUSY breaking.
However, the density of vacua increases $\propto \Lambda_{\rm
SUSY}^{2N}$~\cite{Susskind:2004uv}. Therefore, assigning {\it a
priori} equal probability to each vacuum one arrives at a new
meassure of fine tuning, which takes into account the ``entropy''
associated with the density of vacua, $f_{\rm new} \sim M_H^2\,
\Lambda_{\rm SUSY}^{N}.$ Contrary to $f_{\rm lore}$ requirements,
$f_{\rm new}$ clearly favors a large SUSY breaking scale. For
example, for $\Lambda_{\rm SUSY} \sim 10^{10}~{\rm GeV}$, ${\cal
O} (10^{200})$ vacua become available, enough to fine tune both
the cosmological constant and the Higgs mass.  If we live in this
neighborhood of the landscape, collider data would be expected to
point to the SM rather than SUSY. However, one pays a price for
throwing away MSSM, since it provides a potential explanation for
both dark matter~\cite{Goldberg:1983nd} and the LEP results
favoring the unification of the three SM gauge
couplings~\cite{Dimopoulos:1981yj}.

Split SUSY~\cite{Arkani-Hamed:2004fb} is a relatively new variant of
SUSY which may facilitate the required fine tuning and simultaneously
preserves the achievements of the MSSM. In this model the bosonic
superpartners are heavy, while the extra fermions retain TeV-scale
masses thanks to protection by chiral symmetry. Although split SUSY
does not provide a dynamical explanation for the hierarchy problem,
the assumption of a large-scale SUSY breaking leads to important
information on the underlying parameters and on measurable physical
quantities~\cite{Delgado:2005ek}. In particular, analyses of one
loop~\cite{Arvanitaki:2004eu} and two loops~\cite{Giudice:2004tc}
running of the RG equations, show that split SUSY preserves
unification of couplings. Additionally, as in the MSSM, the lightest
supersymmetric particle provides a possible candidate for cold dark
matter~\cite{Pierce:2004mk}.

It is clear that split SUSY opens new territory for model builders,
gauginos have a symmetry that protects their masses, namely the
$R$-symmetry, so building models where scalars are very massive is
quite natural in theories where this symmetry is not broken, for
example in $D$-term breaking models; it can also happen in theories
with extended supersymmetries and there are already several papers
with string inspired models of split SUSY~\cite{Barger:2004sf}. Hence
there is a strong motivation for phenomenological studies, including
implications for collider-based
measurements~\cite{Zhu:2004ei,Hewett:2004nw,Arvanitaki:2005nq},
electric dipole moments~\cite{Arkani-Hamed:2004yi}, Higgs physics and
electroweak symmetry breaking~\cite{Sarkar:2004cs}, and cosmic ray
physics~\cite{Hewett:2004nw,Anchordoqui:2004bd,Illana:2006xg}. The
latter is the main focus of the present study.

An intriguing prediction of split SUSY, which represents a radical
departure from the MSSM, is the longevity of the gluino.  As mentioned
above, in split SUSY the squarks are very massive and so gluino decay
via virtual squarks becomes strongly suppressed, yielding a $\tilde g$
lifetime of the order of~\cite{Arkani-Hamed:2004fb}
\begin{equation}
  \tau_{\tilde g} \simeq 3 \times
10^{-2} \left(\frac{{\rm TeV}}{M_{\tilde g}}\right)^5
  \left(\frac{\Lambda_{\rm SUSY}}{10^{9}~{\rm GeV}}\right)^4~{\rm s}\,\,,
\label{tau0}
\end{equation}
where $M_{\tilde g}$ is the gluino mass. Quasi-stable colorless
\rhadrons. 
(i.e., carrying one unit of $R$-parity)
are expected to be born when such long-live gluinos become confined with
quarks, anti-quarks, and gluons~\cite{Farrar:1978xj}.

Very strong limits on heavy isotope abundance in turn require the
gluino to decay on Gyr time scales~\cite{Smith:1982qu}, leading to
an upper bound for the scale of SUSY breaking ${\cal O}
(10^{13})$~GeV. More restrictive bounds on $\Lambda_{\rm SUSY}$
can be determined from cosmological
considerations~\cite{Arvanitaki:2005fa}. Specifically, gluino
decays would disturb predictions of big bang nucleosynthesis
(BBN), or distort the cosmic microwave background (CMB), or alter
the diffuse gamma ray background. The details depends on both,
$M_{\tilde g}$ and $\tau_{\tilde g}$.  For example, for $1 \alt
M_{\tilde g}/{\rm TeV} \alt 5,$ to avoid altering the abundances
of D and $^6$Li, $\tau_{\tilde g} \alt 100$~s, implying
$\Lambda_{\rm SUSY} \alt 10^{10}$~GeV~\cite{Arvanitaki:2005fa}.
The relic abundance of lighter gluinos, $M_{\tilde g} \alt
500~{\rm GeV}$, is constrained by COBE~\cite{Fixsen:1996nj},
WMAP~\cite{Spergel:2006hy} and EGRET
observations~\cite{Sreekumar:1997un}. On the one hand, gluinos
that decay during or after the thermalization epoch can distort
the CMB spectrum~\cite{Hu:1993gc} and so are limited by COBE/WMAP
observations.  On the other hand, gluinos that decay after the
recombination epoch give rise to pions which subsequently decay
into $\gamma$-rays that free-stream to us.  The contribution of
such a decay chain to the diffuse $\gamma$-ray
background~\cite{Kribs:1996ac} is limited by EGRET observations.

\begin{figure}
 \postscript{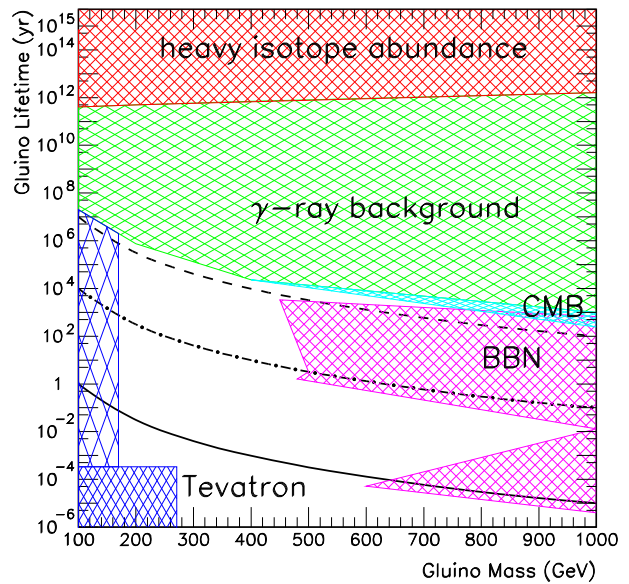}{0.98}
 \caption{Limits on long-lived gluinos. The cross-hatched bands
   indicate excluded regions of the $M_{\tilde g}$-$\tau_{\tilde g}$
   plane from anomalous heavy isotope abundance~\cite{Smith:1982qu},
   CMB~\cite{Fixsen:1996nj} and EGRET~\cite{Sreekumar:1997un}
   observations, BBN predictions~\cite{Arvanitaki:2005fa}, and collider
   data~\cite{Hewett:2004nw,Abazov:2007ht}. Contours of constant values of
   $\Lambda_{\rm SUSY}$ are also
   shown by solid ($10^{10}~{\rm GeV}$), dash-dotted
   ($10^{11}~{\rm GeV}$), and dashed lines ($10^{12}~{\rm GeV}$).}
\label{fig:1}
\end{figure}

Long-lived gluinos are also constrained by collider searches.  Charged
\rhadrons. can be observed as they cross the detector either by their
time delay relative to ultrarelativistic particles~\cite{Abe:1992vr},
or by their anomalously high ionization energy
loss~\cite{Acosta:2002ju}. Besides the energy deposition of
neutral \rhadrons. in the calorimeter is rather soft, and so when
they are produced in association with a high-$p_T$ jet they can be
observed in the monojet channel + missing energy $\met$: CDF Run I
data~\cite{Acosta:2003tz} found a bound of $M_{\tilde g} > 170~{\rm
GeV}$~\cite{Hewett:2004nw}. In addition, \rhadrons. can become
stopped gluinos by losing all of their momentum and coming to rest in
the calorimeter~\cite{Arvanitaki:2005nq}. The D\O\
Collaboration~\cite{Abazov:2007ht} has recently searched for stopped gluinos
decaying into a single jet and a neutralino.  The non-observation of
monojets (above the expected background from cosmic-muon induced
showers) in Run II data implies $M_{\tilde g} > 270~{\rm GeV}$ for
$\tau_{\tilde g} < 3~{\rm hr}.$ All these limits are shown in
Fig.~\ref{fig:1}.  As we will show here, the study of hadronized
gluinos originating in distant astrophysical sources, provide a viable
experimental handle in the region $300 \alt M_{\tilde g}/{\rm GeV}
\alt 500$ - $10^2 \alt \tau_{\tilde g}/{\rm yr} \alt 10^{5}$, which
is yet unexplored.

The main goal of this paper is to describe a full-blown Monte Carlo
simulation of $R$ air showers, and uncover observables which may be
exploited by new experiments like the Pierre Auger
Observatory~\cite{Abraham:2004dt}.  This analysis expands on previous
work~\cite{Gonzalez:2005bc} by including all possible $R$ interactions
and analyzing in detail the potential of the surface array. Before
describing the simulation, we introduce in the following section the
main properties of \rhadron. interactions

\section{Cosmic $\bm{R}$'\lowercase{s}}
\label{2}

The origin cosmic rays is still an open question, with the degree
of uncertainty increasing with rising
energy~\cite{Anchordoqui:2002hs}. Theoretically, one expects the
cosmic ray spectrum to fall off somewhat above $10^{10.7}~{\rm
GeV},$ because the particle's energy gets degraded through
interactions with the cosmic microwave (protons and nuclei) and
radio (photons) backgrounds, a phenomenon known as the
Greisen-Zatsepin-Kuzmin (GZK) cutoff~\cite{Greisen:1966jv}. The
most recent data from the Pierre Auger Observatory in fact do not
indicate (yet) any excess beyond the expected
cutoff~\cite{Roth:2007in}. Because of the rapid energy
degradation, the maximum attainable energy in far away sources can
be considerably higher than the GZK-threshold. However,
measurements of the GeV diffuse gamma ray flux significantly
constrain the cosmic ray production integrated over redshift, and
consequently limit the maximum energy of these particles.
Specifically, the intermediate state of the reaction $p
\gamma_{\rm CMB} \to N\pi$ is dominated by the $\Delta^+$
resonance (because the neutron decay length is smaller than the
nucleon mean free path on the relic photons). Hence, there is
roughly an equal number of $\pi^+$ and $\pi^0.$ Gamma rays,
produced via $\pi^0$ decay, subsequently cascade
electromagnetically on the cosmic radiation fields through $e^+
e^-$ production followed by inverse Compton scattering. The net
result is a pile up of gamma rays at GeV energies, just below the
threshold for further pair production. Therefore, if the
distribution of cosmic ray sources is homogeneous and each source
is characterized by a hard injection spectrum $\propto E^{-1},$
then EGRET measurements in the 100~MeV - 100~GeV
region~\cite{Sreekumar:1997un} limit the maximum proton energy
$\sim 10^{13.5}~{\rm GeV}$~\cite{Semikoz:2003wv}. Since Fermi's
acceleration mechanism predicts a rather steeper spectrum $\propto
E^{-2}$~\cite{Anchordoqui:2002hs}, one can assume a maximum proton
energy $E_{p, {\rm max}}^{\rm lab} \sim 10^{13.7}~{\rm GeV}.$

Gluinos are flavor singlets of a color $SU(3)$ octect that interact
strongly with the octect of gluons and can combine with quarks,
anti-quarks, and gluons to form colorless
hadrons~\cite{Farrar:1978xj}. The bosonic states, $\tilde{g}qqq,$ are
generically called $R$-baryons, whereas the fermionic sates, $\tilde g
q \bar q$ and $\tilde g g,$ are called $R$-mesons and $R$-glueballs,
respectively. Very little is certain about the spectroscopy of these
strongly interacting particles. The most relevant feature is (perhaps)
the difference in mass between $R$-mesons (-glueballs) and
$R$-baryons, because if $M_{R_{m (g)}} + m_N > M_{R_b} + m_\pi$, then
there are exothermic conversions of $R_{m(g)}$ into $R_b$ as the
\rhadrons. propagate in the atmosphere.

\rhadrons. states should be produced in pairs through $pp$ collisions at
 powerful cosmic ray engines (e.g., protons undergoing acceleration in
 compact jets of relativistic plasma interact with those in the
 surrounding gas). The average energy of the produced $R$ in the
 target system is
\begin{equation}
E_R^{\rm lab}  \simeq \sqrt{\frac{E_p^{\rm lab}}{2m_p}}\,
\,\, E_R^{\rm cm} \,\,,
\label{boost}
\end{equation}
where $E_p^{\rm lab}$ is the energy of the proton undergoing
acceleration at the source, $E_R^{\rm cm} \simeq
\sqrt{\hat s}/2$ is the average $R$ energy in the center-of-mass
(c.m.) of the $pp$ collision, and $\hat s$ is the square of the energy
in the c.m. of the parton-parton collision. Now, by restricting
$R$-production to large c.m. energies (say, $\hat s \ge 16 M_R^2$),
from Eq.~(\ref{boost}) we obtain the maximum energy of cosmic $R$'s,
$E_R^{\rm lab} < 10^{7} M_R.$ Thus, the $R$ spectrum cuts off at lower
energy than the cosmic ray spectrum. Since these particles originate
from cosmological distance $d$, to reach the Earth the gluinos must be
remarkably long lived $\tau_{\tilde g} \agt 100 \,(M_{\tilde
g}/500~{\rm GeV}) \,(d/{\rm Gpc})~{\rm yr}$. Moreover, to avoid
deflections on the extra-galactic magnetic fields and the consequent
energy loss due to pair production and other mechanisms (such as
synchrotron or bremsstrahlung radiation), the \rhadron. has to be
neutral~\cite{Farrar:1996rg}. The overall intensity of \rhadrons. is
constrained by its accompanying pion flux, which decays into
$\gamma$-rays and neutrinos that can be confronted with existing
data~\cite{Sreekumar:1997un,Anchordoqui:2002vb}. Unfortunately, the
expected flux of ultra-relativistic (Lorentz factors $\sim 10^7$)
\rhadrons. is found to be very low (less than 6 particles per km$^2$
per millenium~\cite{Anchordoqui:2004bd}), and so the only experimental
method with potential is observation of their interactions in large
volumes of the Earth's atmosphere.

When a hadronized gluino impinges on a stationary nucleon of the Earth
atmosphere a large number (over 140 when summed over all \rhadrons.)
of scattering proceses are possible~\cite{Kraan:2004tz}. Interactions
of $R$-meson states include: {\it (i)} $2 \to 2$ processes, such as
purely elastic, (e.g. $\tilde g d \bar d + uud \to \tilde g d\bar d +
uud$), charge exchange (e.g. $\tilde g d \bar d + uud \to \tilde g u
\bar d + u dd$), and baryon exchange (e.g. $\tilde g d \bar d + uud
\to \tilde g u dd + u \bar d$); {\it (ii)} $2 \to X$ processes
including normal inelastic scattering (e.g. $\tilde g d \bar d + uud
\to \tilde g u \bar d + udd + d \bar d$) and inelastic scattering with
baryon exchange (e.g. $\tilde g d \bar d + uud \to \tilde g u u d + u
\bar d + d \bar u + d \bar d$). Since the final-state pion is so light,
processes with baryon exchange would be kinematically
favored. However, these processes could be dynamically suppressed
because the exchange of two quarks is required. Interaction of $R$-baryon
states include purely elastic, charged exchange, and normal inelastic
scattering. No baryon exchange is possible because of the negligible
probability for a $\tilde g q q q$ to interact with a pion in the
nucleus. Furthermore, this process would be kinematically strongly
disfavored. Consequently, $R$-mesons can convert into $R$-baryons, but
not vice versa.  Interactions of $R$-glueballs are expected to be
similar to those of $R$-mesons. This is because a $g$ is able to split
into a $q \bar q$ state, suggesting that a $\tilde g g$
interacts like (and mixes with) $\tilde g q \bar q$ states.

To establish which of these processes dominates, aside from a model
describing the target (neutron or proton), the relative couplings of all
the processes must be known. The latter requires the calculation of
the Clebsh-Gordon coefficients of isospin-related processes, and the
evaluation of all adittional dynamical effects for all processes.  To
parametrize our ignorance about QCD interactions, here we will
consider all the relevant processes mentioned above (5 for $R$-mesons,
and 3 for $R$-baryons), assigning them different probabilities
{\it ad hoc} so as to explore the entire parameter space.

Predicting the total cross section of an \rhadron. scattering off a
nucleon is non-trivial. However, because of the high c.m. energies
under consideration in this paper, the cross section can be safely
approximated by the geometrical cross section. Moreover, since the
size of the \rhadron. is roughly the same as the size of the
accompanying hadron system, the total cross section for nucleon
scattering can be approximated by the asymptotic values for the cross
sections for normal hadron scattering off nucleons. Therefore, for
$R$-baryons we take $\sigma_{R_b {\rm -} p} (\sqrt{s} \simeq
10^{5}~{\rm GeV}) \approx 140~{\rm mb}$~\cite{Block:2000pg}. This
corresponds to a cross section for scattering off air molecules
$\sigma_{R_b {\rm -air}} (\sqrt{s} \simeq 10^{5})~{\rm GeV} \approx
520~{\rm mb},$ yielding a mean free path in the atmosphere
$\lambda_{R_b} = m_{\rm air}/\sigma_{R_b {\rm -air}} \approx 47~{\rm
g}/{\rm cm}^2$, where we have taken $m_{\rm air} \approx 2.43 \times
10^{-23}~{\rm g}$ (corresponding to an atomic mixture of 78\% N,
21.05\% O, 0.47\% Ar and 0.03\% of other elements). At this
c.m. energy, the $\pi$-air cross section is roughly 90\% of the
$p$-air cross section~\cite{Alvarez-Muniz:2002ne}, hence for $R$-meson
states we set $\lambda_{R_m} = 52~{\rm g}/{\rm cm}^2.$ $R$-glueballs
are expected to have the same cross section as $R$-mesons. This is
because the geometrical cross section is approximated by the high
energy hadron cross section, where gluon exchange dominates (the $gg$
coupling is a factor 9/4 larger than the $qg$ coupling, but a meson
has 2 quarks, resulting in a cross section of a $\tilde g g$ state
which is $(9/4)/(1+1) \simeq 1$ times the cross section for a $\tilde
g q \bar q$ state). With this in mind we set $\lambda_{R_m} \simeq
\lambda_{R_g}.$

In analogy to a billard ball moving through a sea of ping-pong balls,
the $R$ suffers very little energy loss as it traverses the
atmosphere.  Then for $R$-flippers (i.e., $R_{m(g)} \to R_b$) we
assume that the emitted pion has an energy $E_\pi \approx \Gamma
m_{\pi},$ where $\Gamma$ is the Lorentz factor of the incoming
\rhadron.. This means that most of the energy $\sim \Gamma M_{\tilde
g}$ is carried by the accompanying $R_{b}$ produced in the
interaction. Following~\cite{Anchordoqui:2004bd}, for inelastic
collisions we parameterize the fractional energy loss per collision as
$K_{\rm inel} \approx (M_R/{\rm GeV})^{-1}.$ For completness, a
derivation of this relation is given in the Appendix.

\section{The Pierre Auger Observatory}
\label{auger}

The Pierre Auger Observatory (or simply ``Auger'')~\cite{Abraham:2004dt}
is designed to study cosmic rays with energies above about
$10^{9}~{\rm GeV}$, with the aim of uncovering their origins and
nature. Such events are too rare to be directly detected, but the
direction, energy, and to some extent the chemical composition of the
primary particle can be inferred from the cascade of secondary
particles induced when the primary impinges on the upper atmosphere.
These cascades, or air showers, have been studied by measuring the
nitrogen fluorescence they produce in the atmosphere or by directly
sampling shower particles at ground level. Auger is a hybrid detector,
exploiting both of these well established techniques, by employing an
array of water \v{C}erenkov detectors overlooked by fluorescence
telescopes. On clear moonless nights, air showers are simultaneously
observed by both types of detectors, facilitating powerful
reconstruction methods and control of the systematic errors which have
plagued cosmic ray experiments to date.  

The observatory is now operational on an elevated plane in
Western Argentina and is in the process of growing to its final size
of 3,000~${\mathrm{km}}^2$.  The surface detector (SD) consists of an array
of 1600 water tanks deployed on an hexagonal grid with spacing of
1.5~km. These tanks detect the \v{C}erenkov light produced by shower
particles crossing their $1.2~{\rm m} \times 10~{\rm m}^2$ water
volume, thanks to three 9-inch photo-multipliers. The
fluorescence detector (FD) consists of four ensambles of six telescopes,
each of which has a field of view of $30^\circ$ vertically and
$30^\circ$ horizontally (i.e., $180^\circ$ for each fluorescence
detector site).  The geography of Northern site would accommodate
a larger array (of up to 10,370~km$^2$~\cite{Nitz:2007ur}),
allowing higher sensitivity to the low flux of cosmic $R$'s.

Identifying showers themselves is usually straightforward, as there is
essentially no ``background'' for the detectors, at least above their
energy threshold. In the case of Auger, the threshold for the surface
detector is around $10^{8.6}$~GeV, below which less than 10\% of the
showers can trigger three tanks or more, as required. However, full
detection efficiency (i.e., 100\% or ``saturated acceptance'') is
achieved only around $10^{9.5}$~GeV for showers with zenith angle
lower than $60^\circ,$ and lower energy showers are usually discarded
to avoid any complication caused by the energy dependence of both the
detection efficiency and the energy resolution. For fluorescence
detectors, showers with energies as low $10^8$~GeV can be observed.
However, the corresponding acceptance is relatively low, since the
total intensity of the fluorescence light does not allow detection
from a large distance, and the shower maximum is then usually above
the field of view of the telescopes, which prevents accurate
reconstruction. Like for any fluorescence detector, the acceptance of
the eyes of Auger increases with energy (as bigger showers can be seen
from larger distances) and depends on the atmospheric conditions.
However, a precise determination of the fluorescence detector
acceptance is not crucial, thanks to its hybrid nature, the energy
differential flux (or ``spectrum'') is not obtained from the
fluorescence detector, but from the surface detector whose absolute
acceptance is essentially geometrical above saturation and thus is
controlled within a few percent at most.

Identifying the primary particle species is somewhat more difficult as
one has to search for differences in the shower development, which are
usually relatively small and subject to fluctuations associated with
the stochasticity of the first interactions~\cite{Anchordoqui:2004xb}.
However, as we discuss in the next sections the showers initiated by
\rhadrons. have very distinctive characteristics and can be easily
isolated from background.

\section{Air shower simulations}
\label{3}

The AIRES simulation engine~\cite{AIRESManual} provides full
space-time particle propagation in a realistic environment, taking
into account the characteristics of the atmospheric density profile
(using the standard US atmosphere \cite{atmosfera}), the Earth's
curvature, and the geomagnetic field (calculated for the location of
Auger with an uncertainty of a few
percent~\cite{Cillis:2000ij}).

The following particles are taken into account in the AIRES
simulations: photons, electrons, positrons, muons, pions, kaons, eta
mesons, lambda baryons, nucleons, antinucleons, and nuclei up to
$Z=36$.  Nucleus-nucleus, hadron-nucleus, and photon-nucleus
inelastic collisions with significant cross-sections are taken into
account in the simulation. The hadronic processes are simulated
using different models, accordingly to the energy: high energy
collisions are processed invoking an external package ({\sc sibyll}
2.1~\cite{Fletcher:1994bd} or {\sc
qgsjet}II~\cite{Ostapchenko:2006}), while low energy ones are
processed using an extension of the Hillas splitting algorithm
(EHSA)~\cite{Knapp:2002vs}. The threshold energies separating the low
and high energy regimes used in our simulations are 200 GeV and 80
GeV for the {\sc sibyll} and {\sc
  qgsjet}, respectively.  The EHSA low energy hadronic model
used in AIRES is a very fast procedure, effectively emulating the
major characteristics of low energy hadronic collisions. The model is
adjusted to retrieve similar results as the high energy hadronic model
for energies near the transition thresholds previously mentioned, and
the low energy cross sections are calculated from parameterizations of
experimental data. A complete discussion on the low energy hadronic
models is clearly beyond the scope of this paper. A separate report on
this subject will be published elsewhere~\cite{sjsinpreparation}.

The AIRES program consists of various interacting procedures that
operate on a data set with a variable number of records.  Several data
arrays (or stacks) are defined. Every record within any of these
stacks is a particle entry and represents a physical particle. The
data contained in every record are related to the characteristics of
the corresponding particle. The particles can move inside a volume
within the atmosphere where the shower takes place. This volume is
limited by the ground, the injection surfaces, and by vertical planes
which limit the region of interest. Before starting the simulation all
the stacks are empty. The first action is to add the first stack
entry, which corresponds to the primary particle. Then the stack
processing loop begins. The primary is initially located at the
injection surface, and its downwards direction of motion defines the
shower axis. After the primary's fate has been decided, the
corresponding interaction begins to be processed. The latter generally
involves the creation of new particles which are stored in the empty
stacks and remain waiting to be processed. Particles entries are
removed when one of the following events happen: $(a)$ the energy of
the particle is below the selected cut energy; $(b)$ the particle
reaches ground level; $(c)$ a particle going upwards reaches the
injection surface; $(d)$ a particle with quasi horizontal motion exists
the region of interest. After having scanned all the stacks, it is
checked whether or not there are new particle entries pending
further processing. If the answer is positive, then all the stacks are
scanned once more; otherwise the simulation of the shower is complete.

\begin{table}
  \caption{Interaction probabilities for the possible \rhadron. scattering 
processes.} 
\begin{tabular}{cccccc}
\hline
\hline
Hadron & $P_1$ & $P_2$ & $P_3$ & $P_4$ & $P_5$ \\
\hline
$R_b$ &~~~~$0.1\phantom{0}$~~~~&~~~~$0.1\phantom{0}$~~~~&~~~~$\phantom{0}0
\phantom{.0}$~~~~&~~~~$0.8\phantom{0}$~~~~&~~~~~$0\phantom{.00}$~~~\\
$R_m$ & $0.05$ & $0.05$ & $0.1$ & $P_{4}^{m}\phantom{.}$ &
$P_{5}^{m}\phantom{. }$\\
\hline
\hline
\end{tabular}
\label{table}
\end{table}

\begin{figure*}
\begin{minipage}[t]{0.49\textwidth}
\postscript{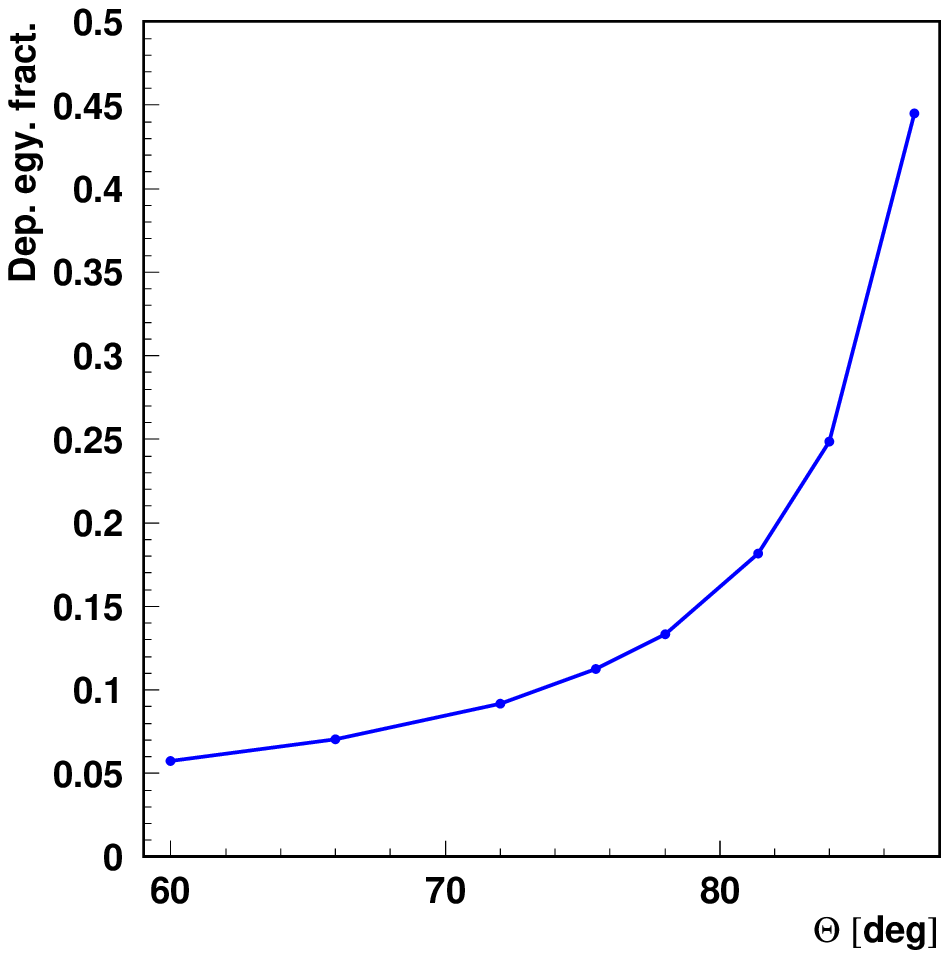}{0.99}
\end{minipage}
\hfill
\begin{minipage}[t]{0.49\textwidth}
\postscript{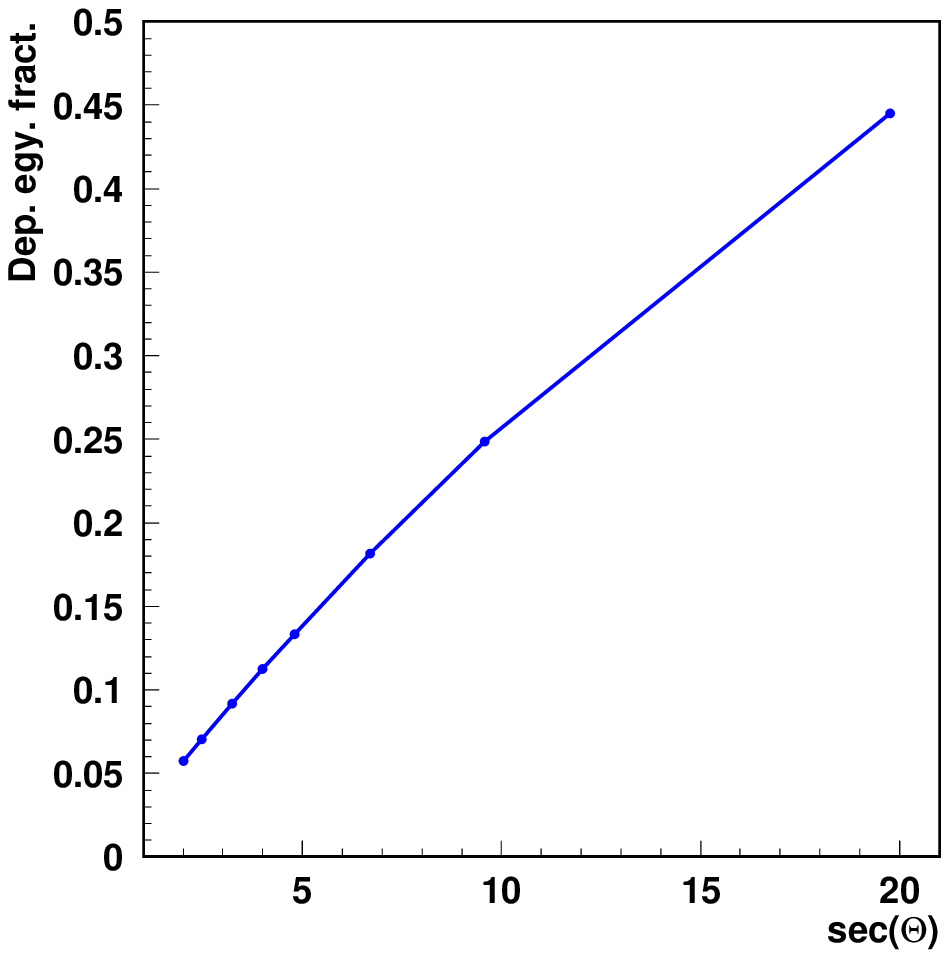}{0.99}
\end{minipage}
\caption{Energy fraction deposited in the atmosphere during \rhadron. air 
showers as a function of the zenith angle $\Theta$ (left) and 
$\sec \Theta$ (right). The curves represent an average over the different 
species. We note that the distinguishing power among the various species is 
very limited for the entire range of $P_4^m$.}
\label{fig:2}
\end{figure*}

AIRES has been successfully used to study several characteristics of
high energy showers, including comparisons between hadronic
models~\cite{Anchordoqui:1998nq}, influence of the LPM
effect~\cite{Cillis:1998hf}, muon bremsstrahlung~\cite{Cillis:2000xc},
and geomagnetic deflections~\cite{Cillis:2000ij} on the shower
development.  AIRES has been also successfully used to determine the
efficiency of  Auger for quasi-horizontal
showers generated by $\tau$-neutrinos~\cite{Bertou:2001vm}, to
estimate the flux of atmospheric muons~\cite{Hansen:2003rq}, and to
study the production of black holes in TeV-scale gravity
models~\cite{Anchordoqui:2001ei}.  For the present analysis, we
prepared a new module to account for the simulation of cosmic $R$'s.
The module includes external parameters such as the type of primary
hadron ($R_b$, or $R_m$), its mass $M_R$, its charge, and its primary
energy $E_R^{\rm lab}$. We adopt the atmospheric mean free path
derived in the previous section.

The total interaction probability is managed by five parameters ($P_i,
\, i = 1,\, \dots,\, 5$) which measure the weights of the different
processes.  (1)~The parameter $P_1$ measures the probability of an
elastic scattering. The program emulates this process by transferring
a small part of the $R$ energy ($\sim 1$~TeV) to an air nucleus which
is injected into the shower. (2)~The parameter $P_2$ measures the
probability of an elastic scattering with charge exchange, in which
there is again a small transfer of energy to an air nucleus, but now
the $R$ primary also flips its charge.  (3)~The parameter $P_3$
accounts for baryon exchange.  In this case, a pion with energy $E_\pi
= m_\pi (\overline {E_R^{\rm lab}} / M_R)$ is injected into the shower
evolution, where $\overline{E_R^{\rm lab}}$ is the energy of the
\rhadron. before the collision. If the incoming \rhadron. is neutral,
then the outgoing pion is charged and vice versa. The \rhadron.
mutates into a baryon with $\widetilde {E_R^{\rm lab}} =
\overline{E_R^{\rm lab}} - E_\pi$.  (4)~In the inelastic process,
controlled by $P_4,$ the \rhadron.  transfers an energy $E_{\rm coll}
= \overline {E_R^{\rm lab}} K_{\rm inel}$ to the shower.  The process
is simulated with the help of the standard packages of AIRES for
nucleon and pion collisions. A neutron (of energy $E_{\rm coll}$)
scatters off an air nucleus if the primary is a $R_b^0$, and a proton
if it is a $R_b^\pm.$ If it is a $R^0_m$ the projectile particle
(simulating the $R$ collision) is a $\pi^0$, whereas if it is a
$R^\pm_m$ the projectile is a $\pi^\pm$. All the secondaries resulting
from this interaction are considered in the subsequent shower
evolution.  Finally, (5)~$P_5$ controls the inelastic scattering with
baryon exchange.  It is simulated as a combination of processes (3)
and (4), i.e, the emission of a pion followed by an inelstic
collision. In our simulations we take, $P_5^m = 0.8 - P_4^m,$ with $0
<P_4^m < 0.8$.

For the simulation engine, the shower starts when the \rhadron.
is added to the previously empty stack.  The injection surface is
located at the top of the atmosphere, spacing the interaction point
according to an exponential distribution with mean equal to
$\lambda_{R_{\rm b (m)}}$. The interaction probabilities $P_i$ are
given in Table~\ref{table}. The \rhadron. is tracked until it reaches
ground level or else its energy is degraded below 100~MeV. New stack
entries are appended to the existing lists for every SM
particle produced in the $R$-interactions. These entries are then
repeatedly processed sequentially by means of the algorithms
implemented in AIRES.

As an illustration, we have run a set of air shower simulations, with
$M_R = 500~{\rm GeV},$ and $E_R^{\rm lab} = 10^{9.7}~{\rm
  GeV}$~\cite{note}. In Fig~\ref{fig:2} we show the energy fraction
dissipated into ``visible'' particles in \rhadron. air showers, as
predicted by our simulations. One can see from the figure that there
needs to be sufficient pathlength for the $R$, with its low
inelasticity, to lose sufficient energy. The experimentally
interesting region to search for \rhadrons. is then $70^\circ \alt
\Theta \alt 90^\circ.$

Because of the very low inelasticity of $R$-air interactions the
leading particle retains most of its energy all the way to the ground,
while the secondary particles promptly cascade to low energies as for
any other air shower.  This results in an ensemble of mini-showers
strung along the trajectory of the leading particle.  Since the
typical distance between mini-showers is about 10 times smaller than
the extent of a single longitudinal profile, it is not possible to
resolve the individual showers experimentally. Instead one observes a
smooth envelope encompassing all the mini-showers, which extends from
the first interaction all the way to the ground, see Fig.~1 in
Ref.~\cite{Gonzalez:2005bc}. The \rhadron. air showers then present a
distinct profile: the flatness of the longitudinal development is
unique to the extremely low inelasticity of the scattering, and can be
easily isolated from background.  However, it turns out that there is
a sharp cutoff in the production of cosmic $R$'s at $E_R^{\rm lab}
\approx 10^{9.5}~{\rm GeV}$~\cite{Anchordoqui:2004bd}, which
unfortunately leads to showers below detection threshold for the
fluorescence method (except for a very small aperture comprised of
regions close to the telescope).

\begin{figure}
 \postscript{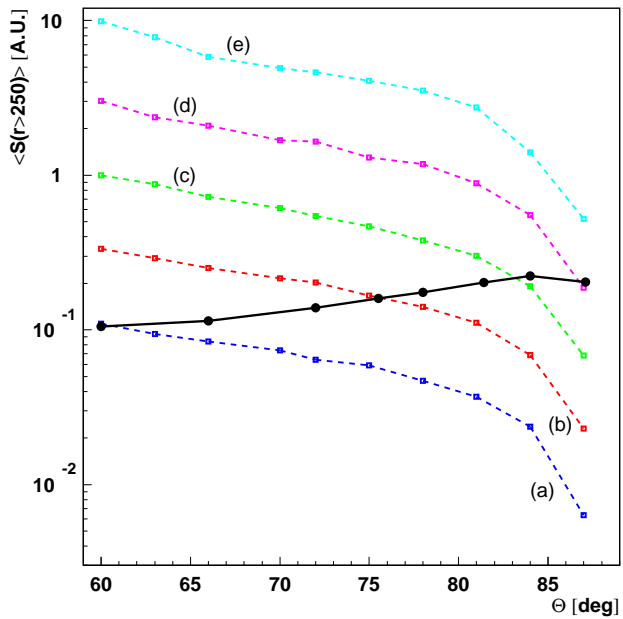}{0.98}
 \caption{Average total signal beyond 250~m from the shower
     core versus shower zenith angle. Signals are normalized to 1 for
     the case of $10^{10}$~GeV proton showers inclined $60^\circ$. The
     dashed lines correspond to proton showers of primary energy: (a)
     $10^{9.0}$~GeV, (b) $10^{9.5}$~GeV, (c) $10^{10.0}$~GeV, (d)
     $10^{10.5}$~GeV, (e) $10^{11.0}$~GeV. The solid line corresponds to
     \rhadron. showers. The lines are only to guide the eye.}
\label{fig:3}
\end{figure}

In Fig.~\ref{fig:3} we show a comparison of the total signal at ground
level for $R$ and proton air showers. To compute the total signal of a
single shower, we first consider all particles reaching the ground
with a distance to the shower axis $r > r_0$, with $r_0$ conveniently
chosen to be 250 m, and for each particle we simulate the \v{C}erenkov
detector response. The total signal of a given shower, $S(r>r_0)$, is
the sum of each particle's individual signal normalized to 1 for
proton showers of $10^{10}$~GeV and incident zenith angle of
$60^\circ$. It is clear that the total $R$ signal at ground level
increases with zenith angle, because of the larger slant-depth.  This
is in sharp contrast to proton showers, in which the signal is reduced
with increasing $\Theta$ because of the greater shower age.

\begin{table}
  \caption{Primary proton energy $E_p^{\rm lab}$ required to produce the same 
total signal at ground level than a \rhadron. with $E_R^{\rm lab} = 
10^{9.7}~{\rm GeV}$, at different zenith angles.} 
\begin{tabular}{c|cccccccc}
\hline
\hline
$\Theta$ & $60^\circ$ & $66^\circ$ & $72^\circ$ & 
$75.5^\circ$ & $78^\circ$ & $81.4^\circ$ & $84^\circ$ & $87.1^\circ$ \\
$E_p^{\rm lab}$~(GeV) & $10^{9.0}$ & $10^{9.2}$ & $10^{9.3}$ & $10^{9.5}$ &
$10^{9.6}$ & $10^{9.8}$ & $10^{10.1}$ & $10^{10.5}$\\
\hline
\hline
\end{tabular}
\label{table2}
\end{table}

As discuss in Sec.~\ref{auger}, the relation between the signal
observed at the surface detectors and the primary energy is determined
using hybrid events in which the fluorescence eyes are thought to
provide a reliable measurement of the total energy. For proton showers
the surface detectors sample about 1\% to 10\% of the shower energy.
Because of the electromagnetic component recycling, the \rhadron.
produces a somewhat larger signal at ground level than one would expect
from standard baryonic showers. As one can check in Table~\ref{2}, for
large zenith angles if one assumes the shower properties are the
characteristics of proton showers then the total primary energy would
be overestimated. Note that this aspect is not compensated by the
calibration procedure, because the $R$-component does not deposit
significant energy in the region of the atmosphere used in the
fluorescence-based calibration. In summary, although the total
contribution to the shower energy is small, the \rhadron. deposits a
disproportional large fraction of their energy close to the ground.
Consequently, cosmic $R$'s would induce a significant signal in the surface
array but not in the fluorescence eyes. In what follows we use these
shower characteristics to construct observables which may be used to
distinguish \rhadron. from traditional cosmic ray showers.

\section{Ground array signal}

\begin{figure}
\begin{center}
\postscript{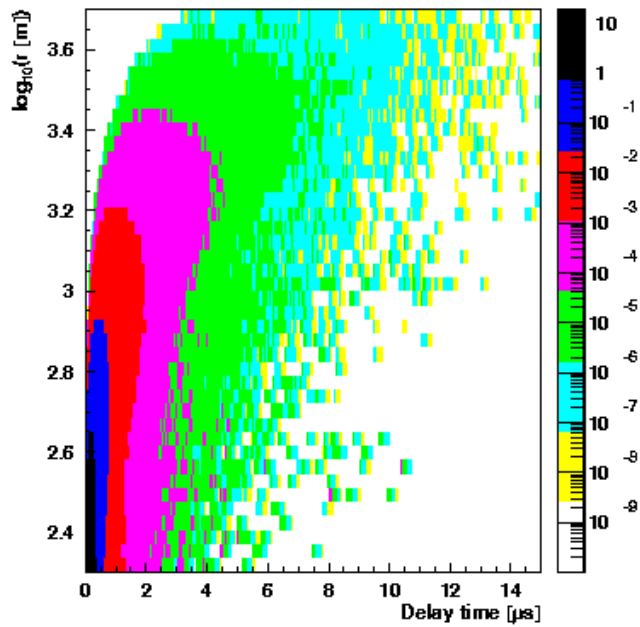}{0.99}
\end{center}
\caption{Lateral vs. arrival time delay distribution for a vertical
  shower initiated by a $10^{9.5}$ GeV proton.}
\label{fig:latimeprotonvert}
\end{figure}

The surface detectors of the Auger Observatory are capable of
measuring the signal associated to an incoming shower as a function of
time. Since a high energy event triggers many detectors, placed at
different distances from the shower axis, it is possible to
reconstruct the lateral-time distribution of the signal $S(r,t)$.
$S(r,t)\,dt$ gives the amount of signal at a (3-dimensional) distance
$r$ from the shower axis, produced at the time interval $[t, t+dt]$.
For convenience, the origin of times is defined for each point in the
ground surface as the instant where a plane orthogonal to the shower
axis, sinchronized with the primary particle and moving towards the
ground at the speed of light, intersects the corresponding point. In
this way, $S(r,t)$ is necessarily zero for negative times. With this
definition, the time $t$ is frequently called ``arrival time delay.''

The total signal at a given distance from the shower axis is the
signal accumulated over all times, that is,
\begin{equation}
S_{\rm tot}(r) = \int_0^\infty S(r,t)\,dt \,\, .
\end{equation}
Other quantities that are usually used in the analysis of SD signals
are:
\begin{itemize}
\item Shower front arrival time, $t_0$ ($t_0\ge0$). This is the time
  corresponding to the first nonzero shower signal at the given
  point. $t_0$ is directly related to the shower front curvature.
\item Partial rise times, $t_x$, defined as the time elapsed until the
  accumulated signal is a fraction $x$ of the total signal, that is,
\begin{equation}
\int_0^{t_x} S(r,t)\,dt = x S_{\rm tot}(r) .
\end{equation}
Common values of $x$ are: 10\%, 50\%, and 90\%. $t_x$ is a growing
function of $r$, especially far from the shower axis.
\end{itemize}

The lateral-time distribution of the signal is a SD observable capable of
characterizing showers initiated by cosmic rays. Consider, for
example a typical shower initiated by a vertical proton. In
Fig.~\ref{fig:latimeprotonvert} the corresponding lateral-time signal
distribution is displayed using a false color (or grayscale) diagram.
From this figure it is possible to notice the main features of such a
distribution: (i) most of the shower particles
arrive near the shower axis, that is, the {\em signal lateral
  distribution\/} $S_{\rm tot}(r)$ decreases with $r$; (ii) $t_0(r)$ 
increases with $r$, as expected, because particles
must travel longer distances and undergo more interactions to reach
positions located far from the shower axis; (iii) the time interval of
the signal at a given point grows with $r$. In the example of
Fig.~\ref{fig:latimeprotonvert} it goes from some 4~$\mu$s at
$r=300$~m to about 12~$\mu$s for $r> 3000$~m.

\begin{figure}
\begin{center}
\postscript{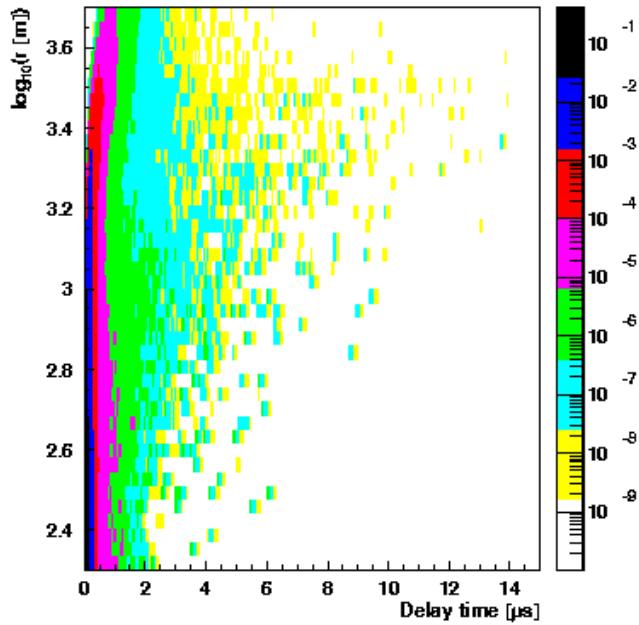}{0.99}
\end{center}
\caption{Same as figure \ref{fig:latimeprotonvert} but in the case of
  $10^{9.5}$ GeV protons inclined 75 degrees.}
\label{fig:latimeproton}
\end{figure}

If the inclination of a shower is increased, the thickness of the air
layer placed between the point where the cosmic particle enters the
atmosphere and ground level, also increases. As a result, the age of
the detected shower increases too. In the case of showers initiated by
hadronic primaries like protons and nuclei, the aging of inclined
showers at ground becomes evident for inclinations larger than 65
degrees, because of the practically complete attenuation of the
electromagnetic component of the shower. For such inclinations, the
muonic component becomes very important (see Fig.~2 of
Ref.~\cite{Cillis:2000ij}), because it produces significant
modifications in the detected signal. In particular, the shower front
becomes flatter, and the signal is concentrated within a relatively
small time span.  These caracteristics show up clearly in
Fig.~\ref{fig:latimeproton}, where $S(r,t)$ is plotted for showers
initiated by protons with the same energy than the showers shown in
Fig.~\ref{fig:latimeprotonvert}, but for an inclination of 75 degrees with
respect to the vertical.

\begin{figure}
\begin{center}
\postscript{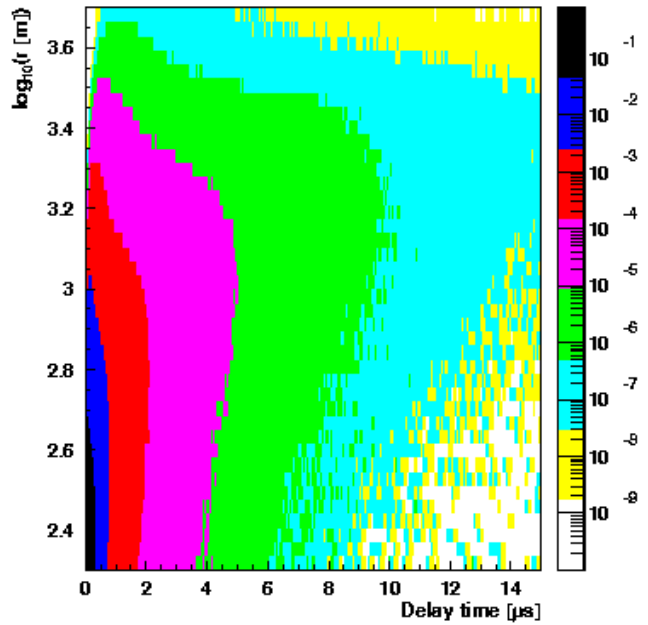}{0.99}
\end{center}
\caption{Same as figure \ref{fig:latimeprotonvert} but for showers
  initiated by 
  $10^{9.7}$ GeV \rhadrons. inclined 75 degrees.}
\label{fig:latimerhadron}
\end{figure}

The lateral-time signal distributions of inclined showers initiated by
\rhadrons. present a substantially different aspect, when compared
with the proton case. In Fig.~\ref{fig:latimerhadron} the signal
distribution corresponding to $10^{9.7}$~GeV $R$-hadron showers
inclined 75 degrees, is displayed. A comparison with the distribution
of Fig.~\ref{fig:latimeproton} leads to the following conclusions: (i)
the \rhadron.  distribution is slighlty more concentrated near the
shower axis, and (ii) the time span of the signal is substantially
larger than in the proton case (note that the primary energy of the
proton showers has been chosen accordingly with Table~\ref{table2}
such that the amount of signal for $r>r_0$ is, on average, the same
for both primaries).

The last feature of the lateral-time distribution of \rhadron. showers
is certainly the most clear signature of such events that could be
found in our simulation study. Combined, in the case of hybrid events,
with a neatly different longitudinal development, and inconsistent
energy measurements, \rhadron. events can be clearly distinguished from
hadronic ones, and also from neutrino initiated showers where the FD
energy determination will be very different from the present case of
\rhadron. showers.

\begin{figure}
\begin{center}
\postscript{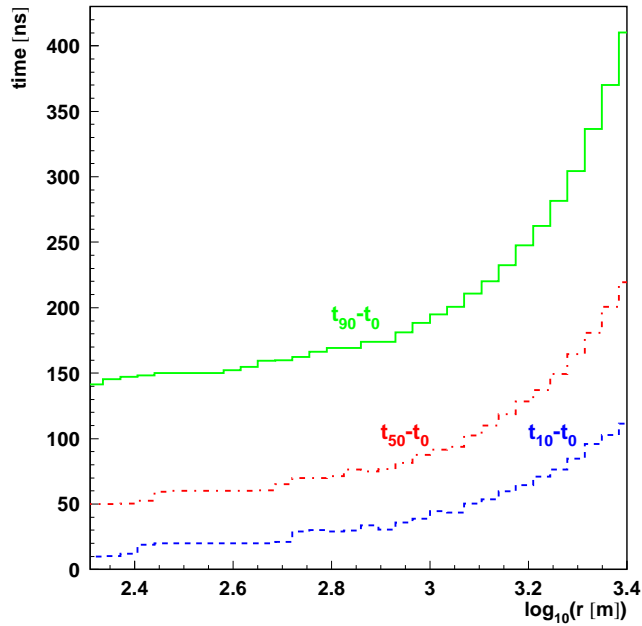}{0.99}
\end{center}
\caption{$t_{10}-t_0$, $t_{50}-t_0$, and $t_{90}-t_0$ plotted versus
  the logarithm of the distance to the shower axis. The data
  corresponds to $10^{9.5}$ GeV protons inclined 75 degrees.}
\label{fig:t1090proton}
\end{figure}

\begin{figure}
\begin{center}
\postscript{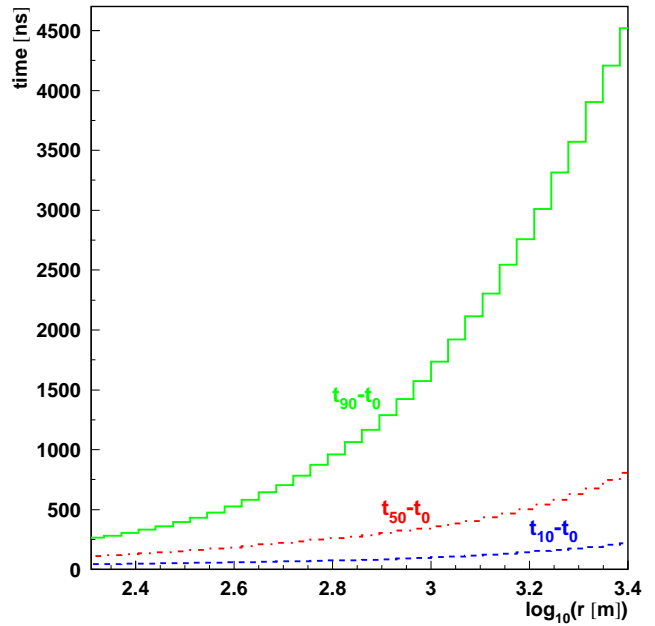}{0.99}
\end{center}
\caption{Same as figure \ref{fig:t1090proton}, but for
  $10^{9.7}$ GeV \rhadrons. inclined 75 degrees.}
\label{fig:t1090rhadron}
\end{figure}

The different time span of signals can be quantified more precisely
studying the observables $t_{10}$, $t_{50}$, and $t_{90}$. Figures
\ref{fig:t1090proton} and \ref{fig:t1090rhadron} contain plots of
these observables as functions of $\log_{10}(r)$, in the case of
proton and \rhadron. showers, respectively. The larger time span of the
signals in the \rhadron. case is evident for all the plotted quantities
(note the different time scales used in either figure).

\begin{figure}
\begin{center}
\postscript{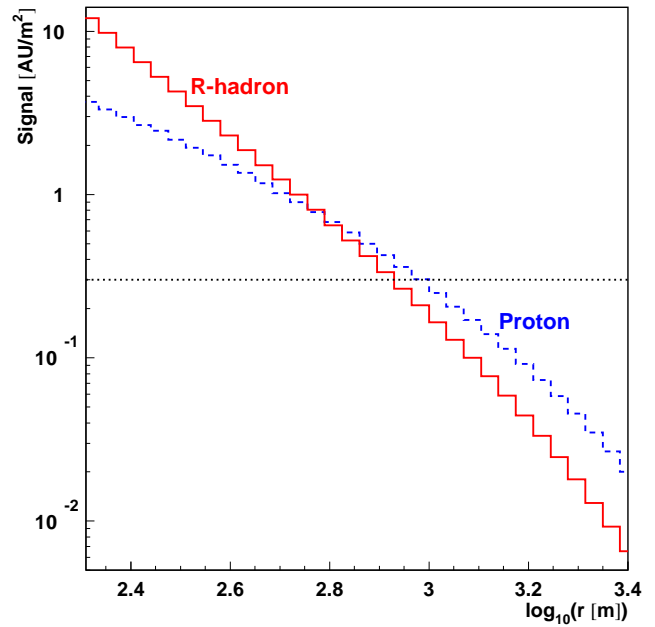}{0.99}
\end{center}
\caption{Water Cherenkov signal plotted versus
  the logarithm of the distance to the shower axis. The solid (dashed)
  histogram correspond to Rhadron (proton) primaries. The horizontal
  dotted line indicates (approximately) the threshold of Auger surface
  detectors.}
\label{fig:latdistprhad}
\end{figure}

The total signal as a function of the distance to the shower axis,
$S_{\rm tot}(r)$, called {\em lateral distribution,\/} is other
fundamental observable that can be measured with the Auger surface
detectors. It is the most important observable for SD energy
determination in the case of showers with an inclination of 
less than 65 degrees.

We have studied the behaviour of the lateral distribution in the case
of \rhadrons., comparing the results with the corresponding ones for
showers initiated with protons. The results, displayed in
Fig.~\ref{fig:latdistprhad}, clearly show that the signal
corresponding to \rhadron. showers is more concentrated near the
shower axis. This also implies a different slope for the distributions
that could eventually be measured. As a reference, a qualitative
indicator of the signal threshold of Auger detectors is also
shown in the figure (dotted line).

\section{Conclusions}

In this paper we have analyzed the characteristics of air showers produced
by gluino-containing hadrons. Using AIRES, we have performed a high
statistics set of full quality showers initiated by \rhadrons.. We
have considered both $R$-baryon and $R$-meson primaries.  The analysis
of standard observables that can be measured by hybrid air shower
experiments like Auger shows that atmospheric cascades initiated by
\rhadrons. are significantly different from ``classical'' showers,
such as for example, those initiated by cosmic protons. Our study
indicates that if cosmic \rhadrons. do exist they would produce a
particular signature that will be visible at Auger: at ground level,
the \rhadron. showers are characterized by the presence of a strong
electromagnetic component at all zenith angles. This implies a much
longer time span for the signal, in comparison with proton showers. At
the same time, the minishowers generated by the passage of the
\rhadron. across the atmosphere produce narrower lateral distributions
than the corresponding ones for the proton case.  

If $R$-hadron events are analyzed with the standard protocol for
hadronic primary showers, a series of inconsistencies will be present.
In particular, the energy determination via ground signal analysis of
very inclined showers~\cite{Facal:2007} will likely lead to a primary
energy overestimation. On the other hand, an eventual hybrid event of
this kind will show a limited, or even below threshold FD signal. This
leads to contradictory FD and SD energy determinations. Moreover, these 
``golden'' events would allow identification of $R$-hadrons from
eventual quasi-horizontal neutrino events that are likely to generate
showers with similar ground signal, but non-negligible fluorescence
contribution~\cite{Bertou:2002}.

The pertinent question at this point is whether existing experiments
have already collected events exhibiting the characteristics of gluino
showers described above. None of the ultrahigh energy cosmic ray
experiments have thus reported such results.  It is interesting to
note, however, that the $~10^{6}~{\rm GeV}$ "Centauro" events detected
at Mt. Chacaltaya~\cite{Lattes:1980wk} might be suggestive of
gluino-induced showers.  In these events, the ratio of hadronic to
electromagnetic components is about 50:1, contrary to the expectation
of dominance of the electromagnetic component in vertical
baryon-induced showers. The most carefully considered explanation to
date is the explosive quark matter model~\cite{Arnison:1982dz}.
Interestingly, though heavy high energy gluinos could also produce
such an inverted hadronic electromagnetic ratio. This is because the
multiple low-inelasticity collisions would result in hadronic
superimposed showers. At detector level ($\sim 5200~{\rm m}$), the
electromagnetic component of the sub-showers would be mostly filtered
out, while the superposed hadronic showers would survive. This is
because the ``low'' energy ($\sim 100$~TeV) electromagnetic
sub-showers induced by high energy $R$-hadrons would develop faster
(being quickly quenched by atmospheric losses) than the high energy
($\sim 10^6$~GeV) electromagnetic subshowers induced by ultrahigh
energy $R$-hadrons.  It is also interesting to note that this
explanation of the Centaruo events does not predict any phenomenon one
might observe at a collider experiment, consistent with the
null-results from UA1~\cite{Arnison:1982dz},
UA5~\cite{Alpgard:1982zs}, and CDF~\cite{Melese:1996ue}.  If in fact,
gluinos are guilty of producing the Centauro events, it would
constitute the first evidence of a finely-tuned universe from a cosmic
ray observation.

\acknowledgments{We would like to thank Haim Goldberg and Carlos Nu\~nez 
for discussions.}

\section*{Appendix}

Consider the process in which two particles of
4-momenta $p_a$ and $p_b$ and masses $M_R$ and $m_N$ scatter two
particles of momenta $p_c$ and $p_d$ and masses $M_R$ and $M_X$, respectively.
Using the total 4-momentum $P$ we define the vector
\begin{equation}
I_\alpha = \epsilon_{\alpha \beta \mu \nu}\,\, P^\beta \,\,p_a^\mu \,\,p_c^\nu
\end{equation}
and write the Lorentz-invariant form
\begin{widetext}
\begin{equation}
I_\alpha \, I^\alpha
=
s\,\,t \,\, (2M_R^2 + m_N^2 + M_X^2  -s -t) - t\,\,
(M_R^2 - m_N^2)\,\, (M_R^2 - M_X^2) -
M_R^2\,\, (M_X^2 - m_N^2)^2
\label{mandel}
\end{equation}
\end{widetext}
in terms of the Mandelstam variables $s = (p_a + p_b)^2 = (p_c +
p_d)^2$ and $t = (p_a - p_c)^2 = (p_b - p_d)^2.$
Note that this squared
invariant when viewed from the c.m. frame reduces to
\begin{eqnarray}
\vec{I}^{*^2} & = & \sqrt{s}\,\, (\vec p_a^{\ *} \times \vec p_c^{\ *})
\nonumber \\
& = & \sqrt{s} \,\, |\vec p_a^{\ *}| \,\, |\vec p_c^{\ *}| \,
\sin \theta^* \,\,,
\end{eqnarray}
where $\theta^*$ is the scattered angle. Consequently, the forward direction
is defined through the condition $I_\alpha I^\alpha = 0$. In the large $s$
limit where
\begin{equation}
s\,\,t \,\, (2M_R^2 + m_N^2 + M_X^2  -s -t) \approx - s^2 \,\,t,
\end{equation}
the minimum momentum transfered can be easily obtained by setting
Eq.~(\ref{mandel}) = 0 and solving for $t_{\rm min}$. All in all,
\begin{eqnarray}
t_{\rm min} & = & - \frac{M_R^2 (M_X^2 - m_N^2)^2}{(M_X^2 -M_R^2)
(m_N^2 - M_R^2) +s^2}
  \nonumber \\
  &  \approx  & - \frac{M_R^2 M_X^4}{s^2} \,\,.
\label{tmin}
\end{eqnarray}
In the c.m. frame, $E_a^* = (s + M_R^2 - m_N^2)/(2 \sqrt{s})$ and
$E_c^* = (s + M_R^2 - M_X^2)/(2 \sqrt{s})$~\cite{Hagiwara:pw}.
Therefore, the invariant quantity $(E_c - E_a)/E_a$ that describes the
inelasticity of the process reads,
\begin{eqnarray}
K_{\rm inel} & \approx &  \frac{(s + M_R^2 - M_X^2) - (s +M_R^2)}{(s+M_R^2)} \nonumber \\
 & \approx & -\frac{M_X^2}{s}\,.
\label{k}
\end{eqnarray}
Now, combining Eqs.~(\ref{tmin}) and (\ref{k}) we obtain
\begin{equation}
K_{\rm inel} \approx \frac{|t_{\rm min}|^{1/2}}{M_R}\,\,.
\label{K}
\end{equation}
The QCD cross section falls off very rapidly and gets negligible for
$t > \Lambda_{\rm QCD}.$ Thus, taking $\Lambda_{\rm QCD} \approx
1~{\rm GeV}$ Eq.~(\ref{K}) leads to $K_{\rm inel} \approx (M_R/{\rm
  GeV})^{-1}.$


\begin{thebibliography}{99}
\bibitem{note} Note that because of the small inelasticity, the main
  characteristics of $R$-hadron air showers are independent of the
  external package used to process the high energy hadronic collisions.
\end{thebibliography}
\end{document}